# DoubleHelix: Structured Cross-Modal Fusion for Audio-Visual Speech Recognition with LLMs

Ziwei Cheng
Software College of Northeastern University
Shenyang, Liaoning, China
chengzw@stumail.neu.edu.cn

Zhenhua Tan✉
Software College of Northeastern University
Shenyang, Liaoning, China
tanzh@mail.neu.edu.cn

Zhuomin Zhu
Software College of Northeastern University
Shenyang, Liaoning, China
zhuzzm@stumail.neu.edu.cn

## Abstract

Audio-visual speech recognition (AVSR) relies on effective fusion of audio and visual modalities, yet existing approaches treat cross-modal interaction as a single-step operation without structured iterative refinement. We present DoubleHelix, a multimodal fusion framework that reformulates fusion as an iterative cross-modal interaction process with adaptive degradation-aware enhancement. The framework comprises three components including ReverseParallelHelix for multi-turn structured interaction with learned alignment constraints, QualitySensor for learning degradation-aware gating signals, and HelixReplication for consistency-guided conditional feature enhancement. Experiments on LRS3 demonstrate that DoubleHelix achieves 0.68% WER on clean audio, outperforming previous best results by 5.6% relative improvement under matched backbone settings. Comprehensive ablation studies validate each component contribution, including targeted analysis of design choices such as asymmetric pathway weighting. The framework shows improved robustness under evaluated babble-noise conditions, achieving 11.6% WER at SNR -5dB.

## 1 Introduction

Audio-Visual Speech Recognition (AVSR) combines audio with lip movement visual information to improve speech recognition accuracy and robustness [20]. Research has explored various fusion strategies including concatenation [26], cross-attention [17, 19], conformer integration [2, 28]. Knowledge distillation from speech foundation models [4, 31] further improves encoder quality. These methods help AVSR systems handle challenging conditions including frame loss [8], visual occlusion [27], and background noise [12, 13]. Cross-modal translation approaches like Lip2Vec [10] and MixSpeech [6] leverage one modality to improve another. A typical AVSR framework has a front-end encoder for feature extraction using models like Whisper [23] for audio and AV-HuBERT [25] for video, and a back-end network for speech recognition.

The integration of Large Language Models (LLMs) into AVSR marks a paradigm shift. LLaMA-AVSR [3] pioneered this direction as the first work to apply LLMs to audio-visual speech recognition. It uses pre-trained Whisper and AV-HuBERT encoders to generate modality tokens processed by LLaMA for speech inference. Following this encoder-LLM paradigm, subsequent works have advanced the field. AVWhisper [29] combines AV-HuBERT and Whisper encoders without LLM decoders. Whisper-Flamingo [24] integrates visual features into Whisper through gated cross-attention. MMS-LLaMA [30] introduces early audio-visual fusion with speaking rate adaptation. Transducer-Llama [9] enables streaming recognition, while SpeechLLM-XL [18] and E2E-ASR [14] address real-time processing. Beyond end-to-end recognition, AVGER [21] applies LLMs for post-hoc error correction, and ClozeGER [16] propose listen-again paradigms. Multi-modality language models [11, 22] extract features with pre-trained encoders and project them into token sequences for LLMs. Parameter-efficient fine-tuning with LoRA [15] reduces training cost while achieving good performance.

**Problem.** Despite these advances, current fusion methods have a fundamental limitation. They treat multimodal fusion as a single-step operation without iterative refinement. MSHF [26] combines features once. Cross-attention [17, 19] does one round of queries between modalities. Conformer-based fusion [2, 28] integrates modalities through single-pass processing. When audio quality drops due to noise, reliability-based methods [8, 13] only weight modalities differently. They reduce audio contribution but do not fix degraded audio using visual information. This discards useful information instead of repairing it. Even the state-of-the-art MMS-LLaMA [30] uses single-step early fusion without progressive refinement.

**Solution and contribution.** We propose DoubleHelix, a fusion module that enables iterative cross-modal interaction within the LLM-based AVSR framework, as shown in Figure 1. We build upon LLaMA-AVSR [3], the pioneering work that first applied LLMs to AVSR, which serves as our baseline. Our method adopts the same encoder-LLM paradigm (Whisper + AV-HuBERT + LLaMA) introduced by LLaMA-AVSR and refined by MMS-LLaMA [30], but focuses on improving the fusion mechanism itself. A listener might mishear a phoneme, use lip shapes to correct it, and reconcile both signals over multiple cycles [20]. Effective fusion should involve repeated exchanges where representations improve through structured interaction. DoubleHelix reformulates fusion as iterative cross-modal interaction with quality-aware enhancement. It has three components. ReverseParallelHelix enables multi-turn interaction between audio and visual features. Modalities refine each other across $K$ turns using rotation matrices and learned pairing constraints. QualitySensor estimates modality quality and HelixReplication does conditional refinement. When audio quality is low, HelixReplication synthesizes improved features using visual templates. When audio quality is high, original features stay unchanged. The main contributions are as follows.

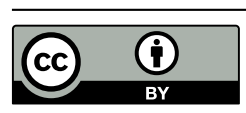

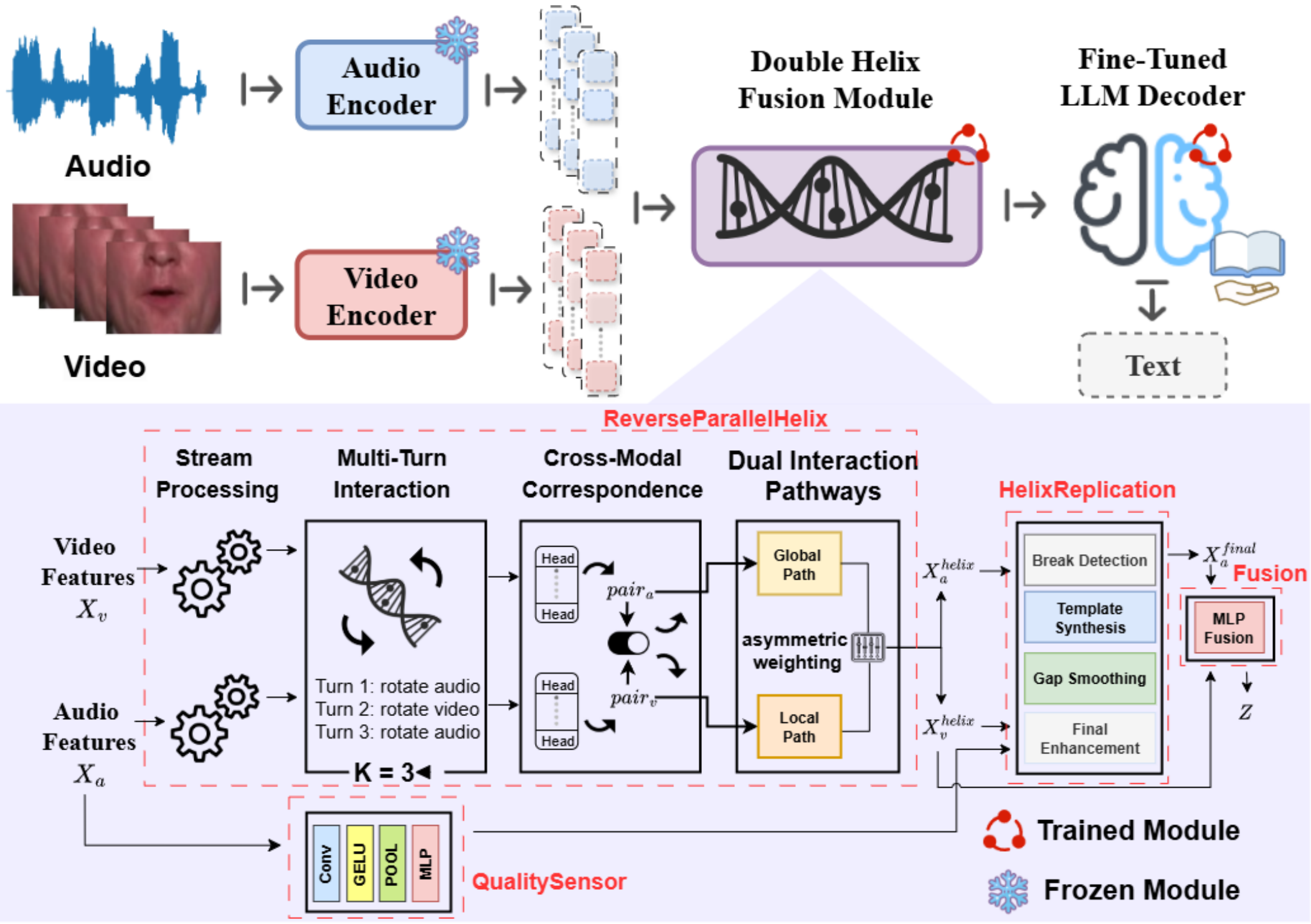

**Figure 1: Architecture overview of DoubleHelix. Audio features from Whisper and visual features from AV-HuBERT are processed through ReverseParallelHelix for iterative cross-modal interaction. QualitySensor estimates audio degradation, gating HelixReplication for adaptive enhancement. The fused representation is then decoded by LLaMA with LoRA fine-tuning.**

- We propose an iterative cross-modal interaction framework. ReverseParallelHelix enables multi-turn representation refinement, turning fusion into a progressive process rather than single-step combination.
- We introduce degradation-aware adaptive enhancement. QualitySensor learns a gating signal from downstream supervision, and HelixReplication enhances degraded audio using visual guidance, moving beyond passive weighting to constructive repair.
- Experiments on LRS3 show DoubleHelix achieves 0.68% WER under clean conditions (5.6% improvement over MMS-LLaMA under matched backbone settings) and shows improved robustness under evaluated babble-noise conditions, achieving 11.6% WER at -5dB SNR.

## 2 Related Work

Audio-Visual Speech Recognition (AVSR) combines audio with visual information from lip movements to improve recognition accuracy and robustness [13, 20]. Research has explored how to leverage audio-visual modality fusion effectively [2, 5, 17, 19, 28]. AVSR systems perform well under frame loss, visual occlusion, and background noise [8, 13, 27]. A typical framework has a front-end for feature extraction and a back-end for speech recognition. Better front-end features improve recognition. Large Language Models (LLMs) enable multi-modality systems that process audio, video, and text together [11, 22].

### 2.1 Audio-Visual Fusion Strategies

Fusion architectures have evolved from concatenation to attention mechanisms. Early approaches use direct feature MSHF [26], which limits cross-modal interaction despite efficiency. Unified-Attention [19] and GILA [17] use cross-attention for bidirectional queries. Unified-Attention [19] uses shared attention alignment, while GILA [17] combines global attention with local alignment. DCIM-AVSR [28] and AVEC [2] modify the Conformer structure.

DCIM-AVSR [28] uses dual Conformer modules, and AVEC [2] handles noise through fusion blocks. These methods perform single-round attention without iterative refinement, which our ReverseParallelHelix addresses. Although knowledge distillation [4, 31] effectively transfers knowledge from speech foundation models, it still use single-step fusion without progressive refinement.

## 2.2 Modality Reliability and Robustness

Handling asymmetric degradation where one modality is more reliable has attracted attention. AV-RelScore [13] train with visual corruption modeling and use reliability scoring to weight modalities. MSRL [5] learn weighting policies through reinforcement learning, while MDA-KD [8] analyze dropout-induced modality bias. These methods adjust contribution weights based on modality quality. They do not enhance degraded representations using intact counterparts, instead down-weighting unreliable modalities and discarding information. PCD [12] improve robustness through contrastive learning, and AVLR [27] handles specific degradation types. Lin and Harte [20] analyze when visual information benefits audio recognition. Lip2Vec [10] maps visual to audio representations, and MixSpeech [6] uses cross-modality self-learning. These approaches leverage one modality to improve another but operate as post-hoc processing. Our QualitySensor provides continuous damage scores that gate enhancement, and HelixReplication synthesizes improved audio features using visual templates when degradation is detected.

## 2.3 LLM-based Audio-Visual Speech Recognition

Aligning pre-trained audio-visual front-ends (Whisper [23], AV-HuBERT [25]) with LLMs enables speech understanding and generation. AVWhisper [29] combines AV-HuBERT and Whisper encoders. Whisper-Flamingo [24] integrates visual features and gated cross-attention into Whisper. LLaMA-AVSR [3] uses pre-trained encoders to generate tokens processed by LLaMA. This method encodes modalities separately before concatenating, potentially missing connections between audio and video. MMS-LLaMA [30] uses early audio-visual fusion that adjusts tokens by speaking rate. Transducer-Llama [9] integrates LLMs into streaming recognition, while SpeechLLM-XL [18] enables real-time processing. E2E-ASR [14] integrates LLMs into first-pass decoding. Beyond end-to-end recognition, generative error correction applies LLMs post-hoc. AVGER [21] encodes audio-visual information to correct N-best hypotheses, and ClozeGER [16] propose listen again paradigms. These post-hoc methods operate on decoder outputs after recognition, whereas our pre-decoding enhancement allows the LLM to use cleaner representations. We adopt the encoder-LLM paradigm from MMS-LLaMA [30] but introduce iterative fusion before the LLM.

## 2.4 Positioning of DoubleHelix

DoubleHelix addresses limitations of prior approaches through three design principles. First, unlike single-step fusion methods (Unified-Attention, DCIM-AVSR), we enable iterative refinement across $K = 3$ interaction turns, achieving plus 10.3 percent improvement over single-turn fusion. Second, unlike reliability weighting that down-weights degraded modalities (AV-RelScore [13]), we enhance degraded representations using visual guidance, achieving plus 31.8 percent improvement over LLaMA-AVSR [3] at -5dB SNR. Third, unlike post-hoc correction that refines decoder outputs (AVGER [21]), we enhance features before decoding, achieving plus 7.2 percent improvement over AVGER [21] at -5dB SNR. Both MMS-LLaMA [30] and DoubleHelix use identical encoders (Whisper-medium + AV-HuBERT large) and decoder (LLaMA-3.2-3B), suggesting iterative fusion as a major contributor under matched backbone settings.

# 3 Method

This paper presents DoubleHelix, an audio-visual speech recognition architecture with structured cross-modal interaction as an iterative refinement process. The proposed method preserves the decoder-only backbone and adopts LoRA for parameter-efficient fine-tuning of the Large Language Model (LLM), while keeping pre-trained audio and visual encoders frozen. To improve robustness under degraded acoustic conditions, DoubleHelix introduces a structured fusion module composed of ReverseParallelHelix for multi-turn cross-modal interaction, QualitySensor for audio quality estimation, and HelixReplication for conditional audio feature repair guided by visual cues. These components form an end-to-end training pipeline, as illustrated in Figure 1.

## 3.1 Notation

In this paper, we denote audio features as $X_a \in \mathbb{R}^{T\times D}$ and visual features as $X_v \in \mathbb{R}^{T\times D}$, where $T$ is the sequence length and $D = 1024$ is the feature dimension. $K$ denotes the number of interaction turns. $H = 8$ is the number of attention heads with head dimension $d_h = D/H = 128$. $\sigma(\cdot)$ denotes the sigmoid function. $[\cdot;\cdot]$ denotes concatenation along the feature dimension. $\langle\cdot,\cdot\rangle$ denotes inner product. $\mathrm{MHA}(Q, K, V)$ denotes multi-head attention with query, key, and value inputs.

## 3.2 Pre-trained Audio–Visual Encoders

The architecture follows an encoder-fusion-LLM paradigm. For audio encoding, we use Whisper-medium [23] (frozen, 24 layers, 1024 dim) to extract acoustic features $X_a$. For video encoding, we use AV-HuBERT large [25] (frozen, 24 layers, 1024 dim) to extract visual features $X_v$. Both encoders stay frozen during training.

## 3.3 ReverseParallelHelix

Conventional fusion does single-step feature combination. ReverseParallelHelix enables multi-turn interaction where modalities refine each other through structured exchange.

**Stream Processing.** Audio and visual features go through convolutional projections with different temporal orientations. Let $\mathrm{LN}(\cdot)$ denote layer normalization.

$$\tilde{X}_a = \mathrm{Conv1d}_{\mathrm{forward}}(\mathrm{LN}(X_a)) \quad (1)$$

$$\tilde{X}_v = \mathrm{Conv1d}_{\mathrm{backward}}(\mathrm{flip}(\mathrm{LN}(X_v))) \quad (2)$$

The forward and backward convolutions process temporal information in complementary directions.

**Multi-Turn Interaction.** Across $K = 3$ interaction turns, representations are refined with learnable rotation matrices. Let $R_k \in$

$\mathbb{R}^{D\times D}$ be the rotation matrix at turn $k$, initialized as orthogonal via QR decomposition. Let $\alpha \in \mathbb{R}$ be a learnable scale initialized to 1.0.

$$X^{(k)} = X^{(k-1)} + \alpha \cdot \mathrm{LN}(X^{(k-1)})R_k \tag{3}$$

Rotations alternate between modalities where turn $k$ rotates one stream while the other anchors, enabling asymmetric refinement.

**Cross-Modal Correspondence.** After rotation refinement, cross-modal attention uses learned pairing matrices. Let $P_h \in \mathbb{R}^{d_h\times d_h}$ be the pairing matrix for head $h$. For each attention head $h$, we compute paired features:

$$\mathrm{pair}_a^{(h)} = X_a^{(h)}P_h, \quad \mathrm{pair}_v^{(h)} = X_v^{(h)}P_h^T \tag{4}$$

where $X_a^{(h)}, X_v^{(h)} \in \mathbb{R}^{T\times d_h}$ are features for head $h$. A learned gate $g^{(h)} \in [0,1]$ modulates pairing strength:

$$g^{(h)} = \sigma\left(\frac{\langle X_a^{(h)}P_h, X_v^{(h)}\rangle}{\sqrt{d_h}} + b\right) \tag{5}$$

where $b \in \mathbb{R}$ is a learnable bias.

**Dual Interaction Pathways.** We implement two pathways for different dependency scales.

$$O_{\mathrm{global}} = \mathrm{MHA}(\mathrm{pair}_a, \mathrm{pair}_v, \mathrm{pair}_v) \tag{6}$$

$$O_{\mathrm{local}} = \mathrm{Conv1d}_{\mathrm{local}}(\mathrm{pair}_v) \tag{7}$$

$O_{\mathrm{global}}$ captures long-range cross-modal dependencies through attention. $O_{\mathrm{local}}$ captures short-range patterns through convolution. Outputs combine with asymmetric weighting to reflect modality importance. We assign higher global-path weight to audio (0.7) and higher local-path weight to visual (0.7), motivated by the observation that audio benefits more from long-range temporal context while visual patterns are more localized.

$$X_a^{\mathrm{helix}} = \mathrm{pair}_a + 0.7 \cdot O_{\mathrm{global}} + 0.3 \cdot O_{\mathrm{local}} \tag{8}$$

$$X_v^{\mathrm{helix}} = \mathrm{pair}_v + 0.3 \cdot O_{\mathrm{global}} + 0.7 \cdot O_{\mathrm{local}} \tag{9}$$

We validate this asymmetric design through ablation in Section 4.4.

## 3.4 QualitySensor

QualitySensor learns a degradation-aware gating signal from audio features for adaptive fusion. Let $d \in [0,1]$ be the damage score indicating audio degradation level.

$$d = \sigma(\mathrm{MLP}(\mathrm{Pool}(\mathrm{GELU}(\mathrm{Conv}(X_a))))) \tag{10}$$

The architecture is Conv1d ($D \to D/4$, kernel 5) followed by GELU, Conv1d ($D/4 \to D/8$, kernel 3), adaptive pooling, MLP, and sigmoid. Unlike methods that require explicit quality labels, our damage score is learned through downstream supervision, where $d$ is trained to correlate with conditions where HelixReplication improves recognition.

A repair weight $w_{\mathrm{repair}} \in [0,1]$ is computed with threshold $\tau = 0.5$ and scale $s = 5.0$,

$$w_{\mathrm{repair}} = \sigma((d-\tau)\cdot s) \tag{11}$$

When $d > \tau$, $w_{\mathrm{repair}}$ approaches 1 and enhancement activates. When $d < \tau$, $w_{\mathrm{repair}}$ approaches 0 and minimal intervention applies. We analyze the correlation between $d$ and noise levels in Section 4.4.

## 3.5 HelixReplication

HelixReplication uses visual features to refine audio representations when audio quality is low.

**Break Detection.** A network identifies temporal regions needing refinement. Let $B \in \mathbb{R}^{T\times 2}$ be the break probability for start and end positions:

$$B = \sigma(\mathrm{MLP}([X_a^{\mathrm{helix}}; X_v^{\mathrm{helix}}])) \tag{12}$$

**Template Synthesis.** Using visual features as template, audio refinements are synthesized:

$$X_a^{\mathrm{new}} = \mathrm{MLP}([X_a^{\mathrm{helix}}; \mathrm{LN}(X_v^{\mathrm{helix}})]) \tag{13}$$

**Gap Smoothing.** Transitions between original and refined features are smoothed. Let $\Delta \in \mathbb{R}^{T\times D}$ be the residual:

$$\Delta = \tanh(\mathrm{MLP}([X_a^{\mathrm{new}} - X_a^{\mathrm{helix}}; X_v^{\mathrm{helix}}])) \tag{14}$$

$$X_a^{\mathrm{repaired}} = X_a^{\mathrm{new}} + \Delta \tag{15}$$

**Final Enhancement.** The final audio representation combines original and repaired features weighted by repair weight:

$$X_a^{\mathrm{final}} = X_a^{\mathrm{helix}} \cdot (1 - w_{\mathrm{repair}}) + X_a^{\mathrm{repaired}} \cdot w_{\mathrm{repair}} \tag{16}$$

## 3.6 Fusion and LLM Integration

Fused representation $Z \in \mathbb{R}^{T\times D}$ combines enhanced audio and visual helix outputs:

$$Z = \mathrm{MLP}_{\mathrm{fusion}}([X_a^{\mathrm{final}}; X_v^{\mathrm{helix}}]) \tag{17}$$

$Z$ is projected to LLM embedding dimension through a linear layer. We use **Llama** as the decoder to generate the final transcription from multimodal inputs $Z$. Specifically, the fused representation output is projected to the LLM embedding dimension and concatenated with the instruction embeddings to construct the multimodal prompt. During training, the target label embeddings are further appended for autoregressive supervision. Therefore, given the multimodal representation $Z$ and the instruction tokens $Text$, the probability of the transcription sequence $Y = \{y_i\}_{i=1}^N$ is defined as

$$p(Y \mid Z, Text) = \prod_{i=1}^{N} p\big(y_i \mid Z, Text, y_{<i}\big), \tag{18}$$

where $y_{<i}$ denotes the tokens generated before step $i$. To adapt the pre-trained decoder efficiently, we adopt **LoRA** [15] for parameter-efficient fine-tuning, where the rank is set to $r = 16$ and the scaling factor is $\alpha = 32$.

## 3.7 Implementation Details

**Rotation Matrix Training.** Each rotation matrix $R_k$ is initialized as an orthogonal matrix via QR decomposition of a random matrix, ensuring stable gradient flow. During training, $R_k$ is jointly optimized with the entire model through standard gradient descent, learning to align cross-modal representations while preserving feature magnitude through the residual connection structure.

**Pairing Matrix Learning.** The per-head pairing matrices $P_h$ are randomly initialized and learned through backpropagation from the cross-modal attention outputs. The gating mechanism $g^{(h)}$ provides adaptive regularization, preventing overfitting to spurious cross-modal correlations. The inner product in the gate computation serves as an implicit alignment objective, encouraging $P_h$ to map audio and visual features to a shared correspondence space.

**QualitySensor Training.** The damage score $d$ is learned through downstream supervision without explicit quality labels. The network learns to produce higher $d$ values for degraded audio where HelixReplication improves recognition, as validated by correlation with noise levels in our experiments. This implicit supervision enables $d$ to serve as a degradation-aware gating signal rather than an absolute quality estimator.

# 4 Experiments

## 4.1 Experimental Setup

**Dataset.** We conduct our experiments primarily on LRS3 [1], a large-scale sentence-level audio-visual speech corpus collected from TED and TEDx videos. The dataset contains approximately **439 hours** of well-aligned speech recordings, covering around 152K utterances from a wide range of speakers and recording environments. Its official validation and test sets provide about 30 hours and 1 hour of carefully annotated audio-visual speech data, respectively. In addition, we make use of VoxCeleb2 [7], a large multilingual audio-visual dataset initially introduced for speaker recognition, with a total duration of 2,442 hours. Note that we retain only the English subset, which contains 1,326 hours of speech video. Since recent multimodal large language model based speech recognition systems are commonly evaluated on LRS3 and VoxCeleb2, we follow this standard experimental setup and use LRS3 as the main benchmark for validating the proposed method.

**Pre-processing**. For visual input preparation, we use the provided facial landmarks to crop a tightly aligned mouth region of interest (ROI) with spatial resolution $96 \times 96$ for each frame. The extracted mouth patches are subsequently converted to grayscale and normalized with global mean and variance statistics, which improves numerical robustness across varying recording conditions. For VoxCeleb2, since manual transcriptions are unavailable, we generate pseudo-labels using the Whisper ASR model and combine the resulting samples with LRS3 for training. After this procedure, the final training corpus contains a total of 1,759 hours of audio-visual speech data.

**Training Setup**. During training, the pre-trained encoders remain frozen, while the fusion module and the LoRA-adapted decoder parameters are optimized. To improve memory efficiency, the LLM is loaded in 4-bit precision. We use AdamW with a peak learning rate of 0.001, weight decay 0.1, and a cosine annealing learning rate schedule. Unless otherwise specified, LoRA is configured with rank $r = 16$ and scaling factor $\alpha = 32$. All models are trained for 7 epochs on 4 NVIDIA A800 80GB GPUs. We report Word Error Rate (WER) as the primary evaluation metric, where lower values indicate better recognition performance.

## 4.2 Main Results: Clean Audio Conditions

**Non-MLLM Methods Analysis.** Among non-MLLM approaches, we observe a performance hierarchy determined by encoder quality and fusion sophistication. Starting from the baseline, Unified-Attention [19] achieves 2.1% WER with standard cross-modal attention. DCIM-AVSR [28] improves to 1.7% WER through dual Conformer modules, a 19% relative improvement from hierarchical fusion. DistillAV [31] advances to 1.3% WER by transferring knowledge from WavLM, showing that encoder pre-training quality impacts performance. AVWhisper [29] achieves the best non-MLLM result at 0.8% WER using Whisper as the audio encoder, a 62% improvement over Unified-Attention [19]. LP Conformer [4] achieves 0.9% WER through specialized Conformer training, close to AVWhisper. All these methods share a limitation in single-step fusion without iterative refinement.

**MLLM-based Methods Analysis.** The introduction of LLM decoders marks a paradigm shift initiated by LLaMA-AVSR [3], the first work to apply LLMs to AVSR. LLaMA-AVSR [3] serves as our baseline, demonstrating the encoder-LLM connection with 0.95% WER using LRS3-only training and 0.77% WER with VoxCeleb2 augmentation, a 19% improvement from additional training data. AVGER [21] achieves 1.45% WER using LLaMA-2-7B with generative error correction, following a different paradigm (post-hoc correction vs end-to-end recognition). MMS-LLaMA [30] advances to 0.72% WER (with VoxCeleb2) using LLaMA-3.2-3B with early fusion, achieving better performance than LLaMA-AVSR's 8B model. This result suggests fusion quality matters more than decoder scale.

**DoubleHelix Results.** Our method achieves state-of-the-art 0.68% WER when trained on LRS3+VoxCeleb2, and 0.83% WER with LRS3-only training. Compared to MMS-LLaMA (0.72% with VoxCeleb2), DoubleHelix achieves a 5.6% relative improvement. Comparing LRS3-only results, DoubleHelix (0.83%) outperforms MMS-LLaMA [30] (0.9%) by 7.8%, demonstrating consistent gains across training configurations. Both DoubleHelix and MMS-LLaMA [30] use identical encoders (Whisper-medium + AV-HuBERT large) and decoder (LLaMA-3.2-3B), isolating iterative fusion as the improvement source. The consistent gap across training setups validates that iterative cross-modal interaction provides fundamental benefits.

## 4.3 Noisy Audio Robustness

To evaluate robustness under acoustic degradation, we test all methods on LRS3 with babble noise at varying SNR levels. Table 2 presents the results with analysis across all conditions.

**Overall Performance.** DoubleHelix achieves the best average WER of 4.3% across all SNR conditions, outperforming LLaMA-AVSR (6.1%) by 29.5% relative improvement and AVGER [21] (4.8%) by 10.4%. Our method achieves the best result at Clean (0.68%), -5dB (11.6%), and 5dB (1.0%), while being competitive at 0dB (3.8% vs AVGER [21] 3.7%), demonstrating consistent robustness across the noise spectrum.

**Per-Condition Analysis.** Examining each SNR level reveals distinct performance patterns.

Clean. DoubleHelix (0.68%) outperforms LLaMA-AVSR [3] (0.77%) by 11.7% and AVGER [21] (1.45%) by 53.1%. The clean-condition improvement validates that iterative fusion benefits recognition even without degradation. Progressive refinement across $K = 3$ turns produces better fused representations than single-step exchange.

-5dB (Extreme Noise). DoubleHelix achieves 11.6% WER, outperforming LLaMA-AVSR [3] (17.0%) by 31.8% relative improvement and AVGER [21] (12.5%) by 7.2%. At this extreme noise level, QualitySensor detects severe degradation (damage score $d$ approximately 0.85), triggering full HelixReplication activation (repair weight $w_{\text{repair}}$ approximately 1.0). LLaMA-AVSR [3] baseline 17.0% WER indicates complete audio failure, while DoubleHelix 11.6% demonstrates effective fallback to visual guidance. 0dB (Moderate Noise). AVGER [21] achieves the best result at 3.7% WER, slightly outperforming DoubleHelix (3.8%) by 2.7%. LLaMA-AVSR [3] reaches 4.4% WER. AVGER advantage at this specific condition comes from its post-hoc correction mechanism. Given moderately corrupted N-best hypotheses, the LLM can leverage semantic knowledge to correct errors. However, this advantage disappears at more extreme conditions. 5dB (Mild Noise). DoubleHelix achieves the best result at 1.0% WER, outperforming LLaMA-AVSR [3] (2.1%) by 52.4% and AVGER [21] (1.7%) by 41.2%. This 52.4% improvement, our largest gain across all conditions, demonstrates that quality-aware enhancement is most effective when audio is moderately degraded but still informative.

**Table 1: Experimental results on LRS3. WER denotes word error rate where lower values indicate higher accuracy. Results shown as X/Y indicate LRS3-only and LRS3+VoxCeleb2 training. The best result is highlighted in bold.**

| Method | Year | Encoder/Decoder | MLLMs | WER |
|---|---|---|---|---|
| Unified-Attention [19] | 2024 | Conformer/Transformer | × | 2.1 |
| LP Conformer [4] | 2024 | Conformer/LSTM | × | 0.9 |
| DCIM-AVSR [28] | 2025 | Conformer/Transformer | × | 1.7 |
| DistillAV [31] | 2025 | WavLM/Transformer | × | 1.3 |
| AVWhisper [29] | 2025 | Whisper/Transformer | × | 0.8 |
| AVGER [21] | 2025 | HuBERT/LLaMA2(7B) | √ | 1.45 |
| LLaMA-AVSR [3] | 2025 | Whisper+AV-HuBERT/LLaMA3.1(8B) | √ | 0.95/0.77 |
| MMS-LLaMA [30] | 2025 | Whisper+AV-HuBERT/LLaMA3.2(3B) | √ | 0.9/0.72 |
| DoubleHelix (Ours) | 2026 | Whisper+AV-HuBERT/LLaMA3.2(3B) | √ | 0.83/**0.68** |

**Table 2: WER (%) on LRS3 with babble noise at various SNR levels. Δ: Performance changes relative to LLaMA-AVSR, where '+' indicates a relative improvement.**

| Method | Clean | -5dB | 0dB | 5dB | Avg |
|---|---|---|---|---|---|
| AVGER [21] | 1.45 | 12.5 | **3.7** | 1.7 | 4.8 |
| LLaMA-AVSR [3] | 0.77 | 17.0 | 4.4 | 2.1 | 6.1 |
| **DoubleHelix** | **0.68** | **11.6** | 3.8 | **1.0** | **4.3** |
| Δ (%) | +11.7 | +31.8 | +13.6 | +52.4 | +29.5 |

**Improvement Scaling Pattern.** The relative improvements over LLaMA-AVSR [3] follow a pattern including 11.7% (Clean), 13.6% (0dB), and 31.8% (-5dB). This monotonic increase validates our quality-aware design. As noise severity increases, QualitySensor triggers stronger enhancement, providing proportionally larger benefits. The non-monotonic pattern at 5dB (52.4%, the largest improvement) indicates that mild degradation presents an optimal operating regime for adaptive enhancement. There is sufficient degradation to benefit from enhancement, but not so severe that recovery is impossible.

**Limitations of Noise Evaluation.** We note that our noise experiments are conducted with babble noise only. While this represents a common and challenging noise type for speech recognition, evaluation across additional noise types (white noise, street noise, music) would provide a more comprehensive robustness assessment. We leave this for future work.

**Table 3: Ablation study on LRS3 at -5dB SNR. Each row removes one component from full DoubleHelix. Δ: Performance changes relative to the full model, where '-' indicates a relative decrease.**

| Configuration | WER (%) | Δ(%) |
|---|---|---|
| DoubleHelix (full) | **11.6** | – |
| *Iterative Interaction Ablations* | | |
| Without ReverseParallelHelix | 14.2 | -22.4 |
| Single turn ($K = 1$) | 12.8 | -10.3 |
| Two turns ($K = 2$) | 12.1 | -4.3 |
| Without Cross-Modal Correspondence | 11.9 | -2.6 |
| Without Global Path | 11.8 | -1.7 |
| Without Local Path | 11.7 | -0.9 |
| Symmetric weighting (0.5/0.5) | 12.0 | -3.4 |
| *Adaptive Enhancement Ablations* | | |
| Without QualitySensor (always enhance) | 12.5 | -7.6 |
| Without QualitySensor (never enhance) | 12.4 | -7.0 |
| Without HelixReplication | 11.8 | -1.7 |

## 4.4 Ablation Study

To understand the contribution of each component, we conduct ablation experiments at -5dB SNR where cross-modal enhancement is most critical. Table 3 presents the results.

**Iterative Interaction is Essential.** Removing the entire ReverseParallelHelix module causes the largest degradation (-22.4%), increasing WER from 11.6% to 14.2%. This confirms that iterative cross-modal interaction is not an incremental improvement but the core contribution. Single-step fusion (equivalent to $K = 1$) degrades by 10.3%, while $K = 2$ degrades by only 4.3%. The progressive improvement from $K = 1$ to $K = 3$ suggests that each interaction turn provides meaningful refinement. Early turns establish coarse alignment, middle turns refine correspondence, and later turns reconcile residual inconsistencies.

**Structured Alignment Mechanisms.** Within ReverseParallelHelix, removing Cross-Modal Correspondence causes -2.6% degradation, while removing Global Path and Local Path causes -1.7% and -0.9% respectively. These results validate our dual-pathway design. Global Path captures long-range dependencies through attention, while Local Path captures short-range patterns through convolution. The larger impact of Global Path suggests that cross-modal alignment benefits more from global context than local patterns.

**Asymmetric Weighting Design.** To validate the 0.7/0.3 asymmetric weighting, we compare against symmetric weighting (0.5/0.5 for both modalities). Symmetric weighting causes -3.4% degradation, confirming that the asymmetric design benefits performance. We hypothesize that audio features benefit more from global temporal context due to the sequential nature of speech, while visual lip movements are more localized in time.

**Adaptive Enhancement Strategy.** The ablation on QualitySensor reveals a critical insight. Both fixed strategies (always-enhance and never-enhance) cause significant degradation (-7.6% and -7.0% respectively). Always-enhance harms performance by applying unnecessary corrections to clean or mildly degraded audio, introducing artifacts. Never-enhance loses robustness benefits by failing to repair severely degraded audio. The comparable degradation of both strategies validates our adaptive gating design. QualitySensor provides the critical ability to distinguish when enhancement is beneficial. HelixReplication alone contributes -1.7%, smaller than the gating mechanism, indicating that when to enhance matters more than how to enhance.

**Component Contribution Ranking.** Summarizing the ablation results, we rank components by contribution. First, iterative interaction ($K = 3$ vs $K = 1$) contributes plus 10.3%. Second, quality-aware gating contributes plus 7.6%. Third, Cross-Modal Correspondence contributes plus 2.6%. Fourth, HelixReplication contributes plus 1.7%. This ranking aligns with our design philosophy where iterative refinement is the primary driver, with adaptive enhancement providing complementary robustness.

## 4.5 Qualitative Analysis

We further examine how DoubleHelix behaves under severe acoustic degradation. Specifically, we visualize intermediate representations and repair signals on samples at SNR = $-5$ dB. Figure 2 shows the audio and visual features of a representative noisy sample in a shared PCA space. Across helix turns, the modality distributions become more structured and their centers move closer, suggesting that *ReverseParallelHelix* progressively improves cross-modal consistency through iterative refinement. Figure 3 presents the break-point activation map produced by *HelixReplication*. The activations are concentrated on localized temporal regions, indicating that the model performs selective repair rather than uniform enhancement over the entire sequence. To assess this behavior more broadly, Figure 4 summarizes the audio-visual feature distance across helix turns over the noisy set. The distance distribution shifts downward from early to later turns, providing evidence that the proposed iterative interaction progressively improves modality alignment under strong noise. Finally, Figure 5 plots the estimated damage score and the corresponding repair weight for noisy samples. Although the repair response is mild in magnitude, it follows the trend of the estimated degradation, indicating that enhancement is modulated by the quality-aware gating mechanism.

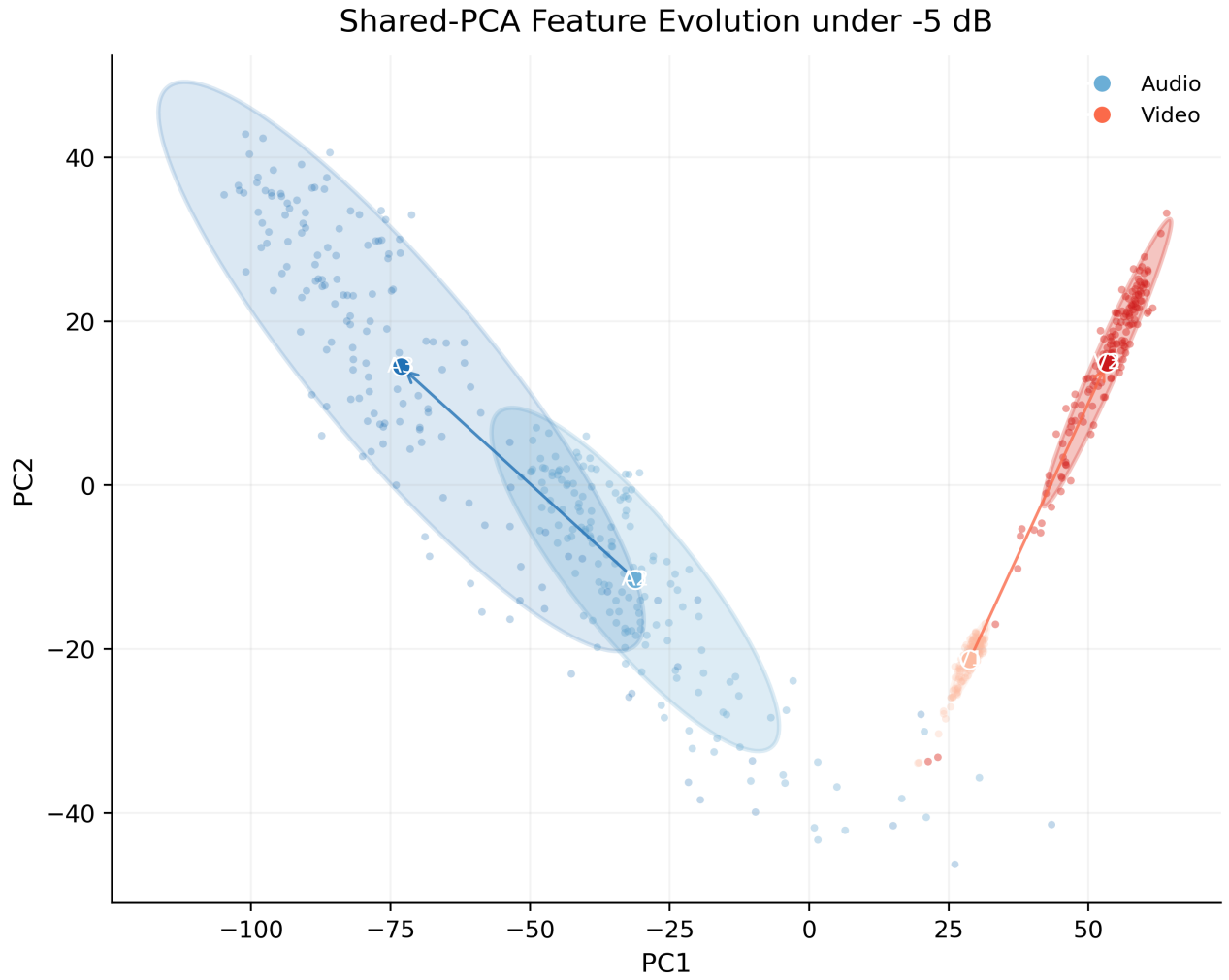

**Figure 2: Shared-PCA feature evolution under severe noise. Audio and visual representations of a representative -5 dB sample are projected into a shared PCA space. The feature distributions become progressively more structured across helix turns, and the modality centers move closer, suggesting improved cross-modal consistency through iterative interaction.**

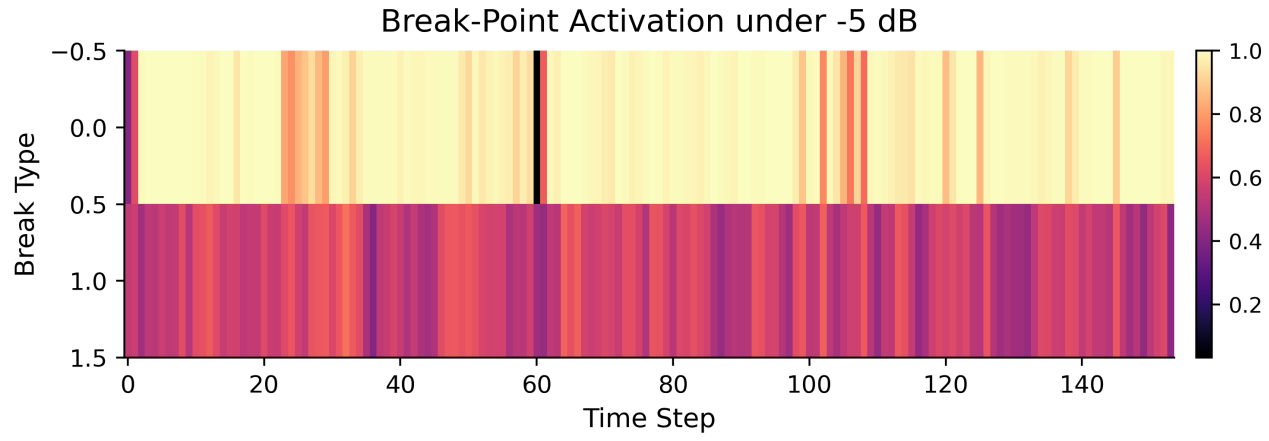

**Figure 3: Break-point activation map produced by HelixReplication. The model selectively activates repair on temporally localized regions instead of enhancing the entire sequence uniformly, indicating conditional repair behavior under strong acoustic degradation.**

## 4.6 Hyperparameter Sensitivity

We analyze sensitivity to two key hyperparameters: the number of interaction turns $K$ and the repair threshold $\tau$. Table 4 presents results at -5dB SNR.

**Interaction Turns $K$.** Performance improves monotonically from $K = 1$ (12.8% WER) to $K = 3$ (11.6% WER), then saturates at $K = 4$ (11.6% WER). This saturation point aligns with theoretical intuition. After 3 to 4 turns, cross-modal alignment has converged and additional turns provide diminishing returns. We select $K = 3$ as the default, balancing performance with computational efficiency (each turn adds approximately 5% overhead).

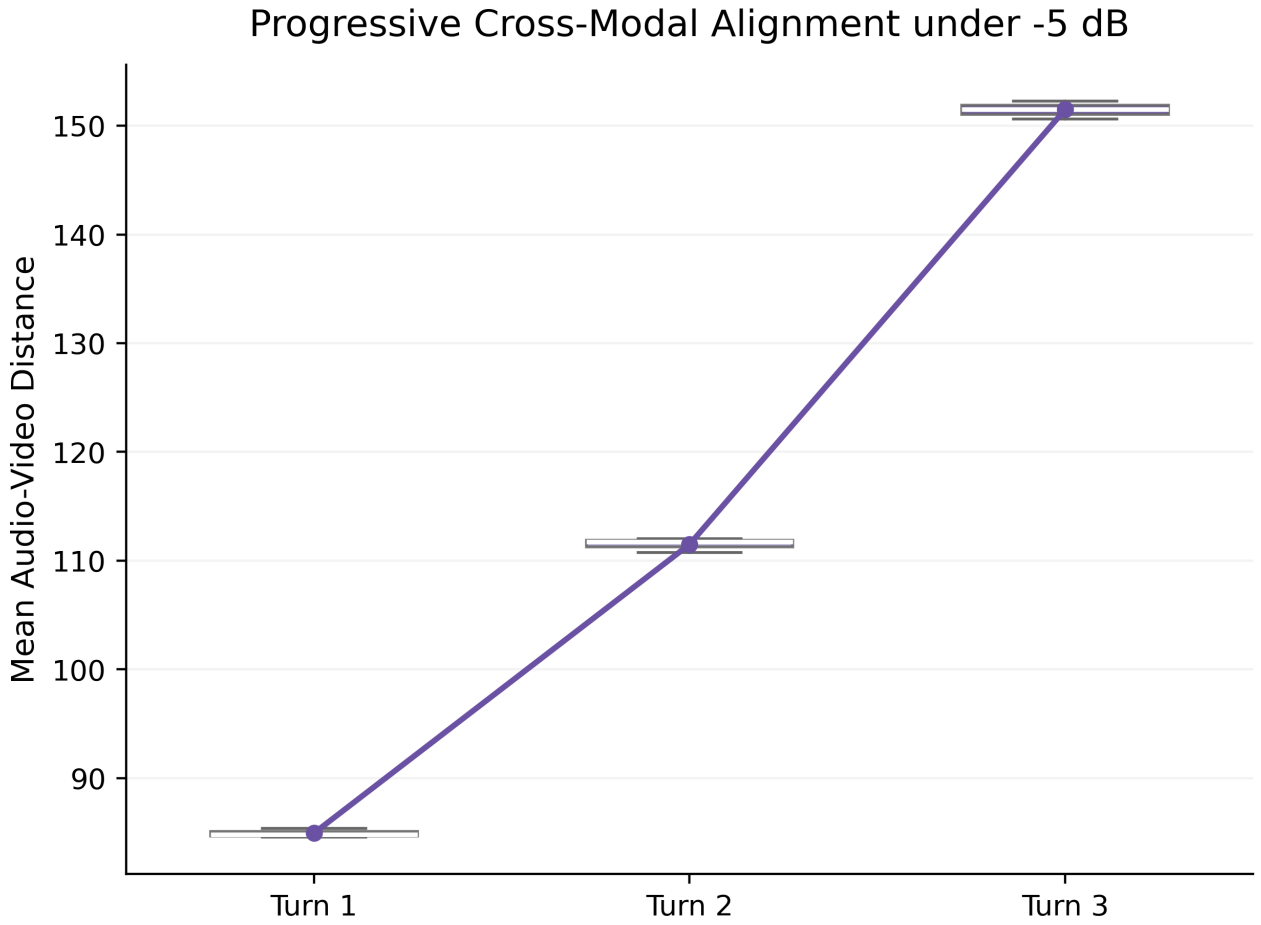

**Figure 4: Cross-modal alignment improvement across helix turns. Boxplots show the distribution of audio-visual feature distance on the -5 dB set. The distance decreases over turns, indicating that ReverseParallelHelix progressively improves alignment under severe noise.**

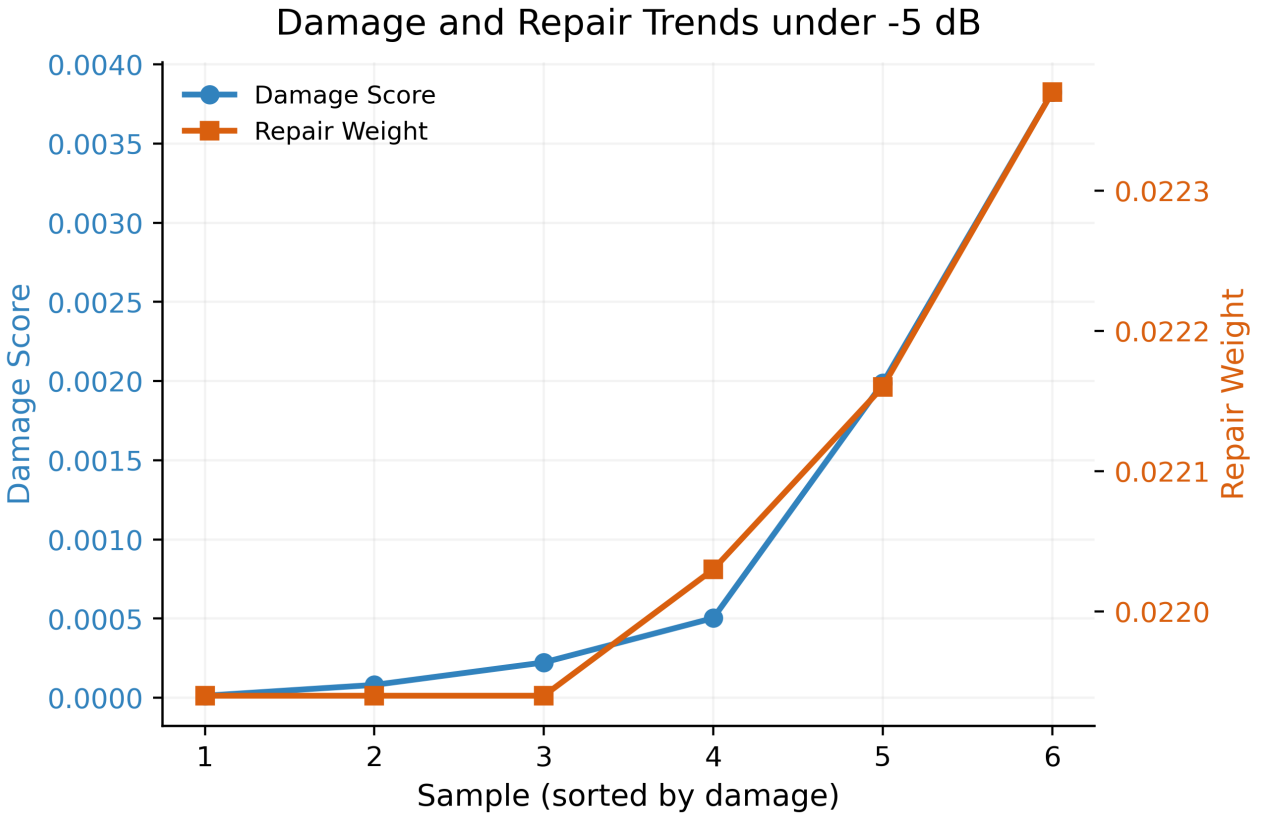

**Figure 5: Damage-aware repair behavior under severe noise. Samples are sorted by estimated damage score. The repair weight follows the variation of the damage estimate, showing that enhancement is modulated by the quality-aware gating mechanism rather than being activated unconditionally.**

**Repair Threshold $\tau$.** The threshold $\tau$ controls the sensitivity of HelixReplication activation. Lower values ($\tau = 0.3$) cause more frequent enhancement, slightly harming performance (-2.6%) due to unnecessary corrections. Higher values ($\tau = 0.7$) delay enhancement, missing opportunities to repair degraded audio (-4.3%). The default $\tau = 0.5$ achieves the best balance, activating enhancement when damage score $d$ exceeds the median of its expected distribution. The relatively small degradation range (-2.6% to -4.3%) indicates that DoubleHelix is robust to threshold choice within a reasonable range [0.4, 0.6].

**Table 4: Hyperparameter sensitivity analysis at -5dB SNR. Δ: Performance changes relative to the full model, where '-' indicates a relative decrease.**

| Configuration | WER (%) | Δ |
|---|---|---|
| Default ($K = 3, \tau = 0.5$) | **11.6** | – |
| *Interaction Turns K* | | |
| $K = 1$ | 12.8 | -10.3 |
| $K = 2$ | 12.1 | -4.3 |
| $K = 4$ | 11.6 | 0 |
| *Repair Threshold $\tau$* | | |
| $\tau = 0.3$ | 11.9 | -2.6% |
| $\tau = 0.7$ | 12.1 | -4.3% |

**Practical Recommendations.** Based on these analyses, we recommend $K = 3$ for general use, with $K = 4$ providing marginal improvement at higher computational cost. We recommend $\tau = 0.5$ as default, with adjustment to $\tau = 0.4$ for cleaner test conditions or $\tau = 0.6$ for consistently noisy environments.

## 5 Conclusion

In this paper, we introduced DoubleHelix, a multimodal fusion framework for audio-visual speech recognition. Unlike conventional one-step fusion methods, DoubleHelix treats fusion as an iterative cross-modal interaction process and further incorporates adaptive degradation-aware enhancement. The framework includes three tightly coupled components. ReverseParallelHelix performs multi-turn structured interaction. QualitySensor learns a degradation-aware gating signal from downstream supervision. HelixReplication conducts consistency-guided conditional enhancement when degradation is detected. Experiments on LRS3 show that DoubleHelix achieves 0.68% WER on clean audio, which corresponds to a 5.6% relative improvement over MMS-LLaMA under matched backbone settings. The model also shows improved robustness under evaluated babble-noise conditions and reaches 11.6% WER at SNR = -5 dB. In addition, ablation results verify the contribution of each component, including targeted analysis of asymmetric pathway weighting and damage score correlation with noise levels. Overall, the results suggest that progressive cross-modal interaction together with degradation-aware enhancement offers an effective direction for multimodal fusion in AVSR.

## Acknowledgments

This work is supported by the National Key Research and Development Program of China under Grant No.2023YFC3306201.

# A Detailed Transcription Case Studies

To qualitatively understand the robustness improvement brought by DoubleHelix under complex conditions, we provide representative transcription examples from the LRS3 test set, as shown in Figure 6. We evaluated the transcription results under clean noise, -5dB, and -15dB. In complex segments with severely degraded acoustic signals, DoubleHelix utilizes its structured iterative interactions to maintain a more stable multimodal representation, thus obtaining accurate transcription results that closely match the true values. Specifically, DoubleHelix can still correctly identify the content at a signal-to-noise ratio of -5dB. At -15dB, DoubleHelix incorrectly identifies "prototype" as "product launch," but most other content is still correctly identified.

To explore the potential for practical applications of DoubleHelix, we occluded the mouth region, as shown in Figure 7. When the mouth region is occluded, the recognition performance of DoubleHelix degrades. Specifically, the areas highlighted in red indicate omissions in content recognition. Analysis reveals that DoubleHelix does not fundamentally address the issue of visual occlusion. Therefore, future work will focus on conducting targeted research on this topic.

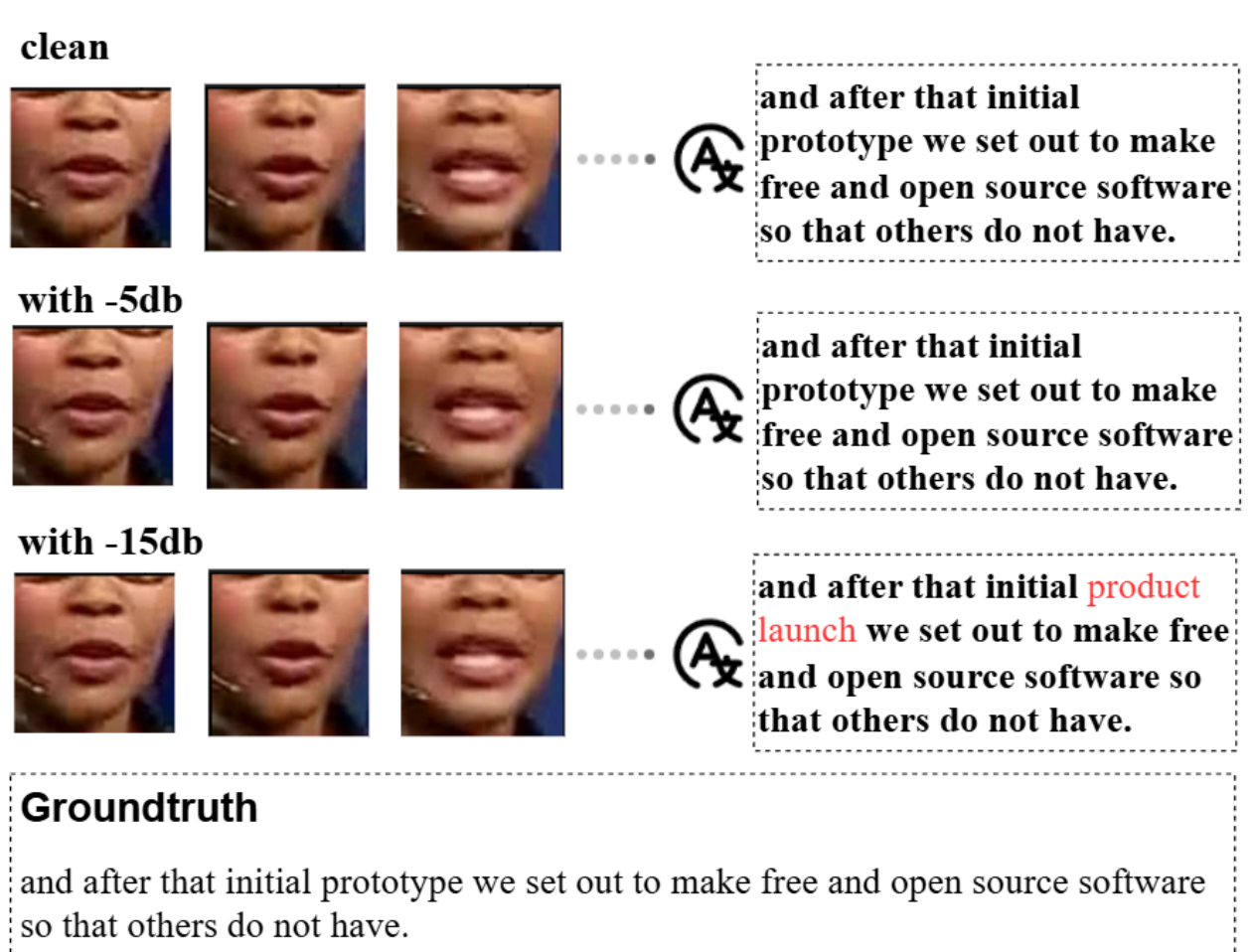

**Figure 6: Examples of transcription under acoustically degraded conditions. Bold text indicates correct recognition. Red text indicates errors.**

# B Computational Complexity and Latency Analysis

A potential concern regarding the DoubleHelix framework is the additional computational overhead introduced by the structured multi-turn interaction. In this section, we provide a detailed breakdown of the parameters and floating-point operations (FLOPs) for the fusion module, as shown in Table 5. The DoubleHelix module introduces a total of 37.25M trainable parameters. While this is larger than a simple early-concatenation baseline, it constitutes less than 1% of the overall model capacity when compared to the frozen backbone encoders (Whisper-medium and AV-HuBERT-large, ~1000M) and the LLaMA-3.2-3B decoder. From a computational perspective, the module requires 34.79 GFLOPs per sequence. Given that the autoregressive decoding of a sequence by a 3B-parameter LLM alone typically demands ~900 GFLOPs, the relative FLOPs overhead introduced by DoubleHelix is marginal (<4% of the total inference computation).

**Table 5: Complexity breakdown of the DoubleHelix fusion module.**

| Component | Params | FLOPS |
|---|---|---|
| QualitySensor | 1.41M | 0.43G |
| ReverseParallelHelix ($K = 3$) | 15.90M | 15.04G |
| HelixReplication | 17.85M | 16.50G |
| **DoubleHelix Total** | 37.25M | 34.79G |

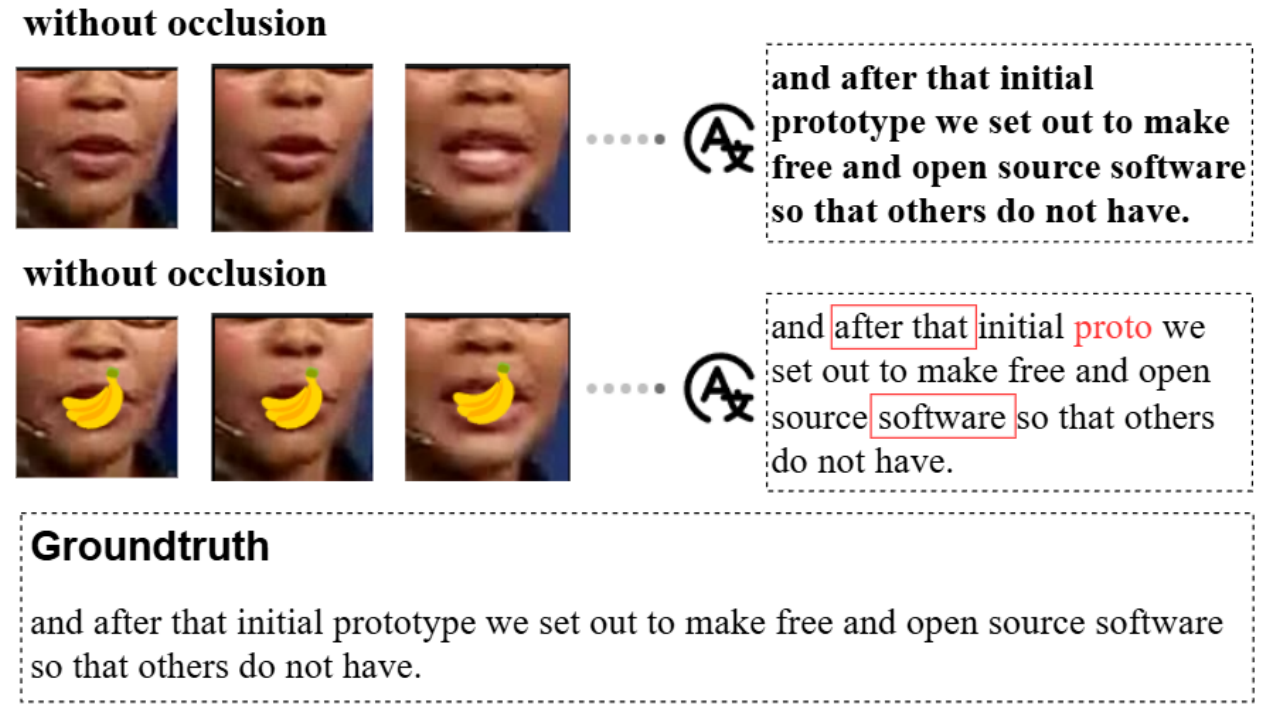

**Figure 7: Representative transcription examples under visual occlusion conditions. Bold text indicates correct recognition. Red text indicates errors. Red boxes indicate omissions.**

# C Theoretical Interpretation of Structured Interaction

We visualize the Shared-PCA feature evolution across helix turns to illustrate the structural changes during fusion, as shown in Figure 2. Notably, the cross-modal distributions do not merely overlap, but rather converge into a more structured and compact dual-cluster topology. Standard cross-modal attention mechanisms (e.g., in DCIM-AVSR) are often susceptible to representation collapse, where in audio and visual features are aggressively projected into a shared manifold to minimize attention or contrastive losses, thereby discarding modality-specific discriminative details. The ReverseParallelHelix avoids this issue via its dual-pathway asymmetric weighting and residual orthogonal rotations. Specifically, the rotation matrices $R_k$ act as a form of implicit orthogonal regularization. Instead of collapsing the modalities into a single point, the structured interaction refines the intra-modal geometry while establishing a stable cross-modal correlation boundary. Consequently, the model learns a structured separation rather than an indiscriminate intersection, preserving fine-grained phonetic and visual cues for the downstream LLM decoder, which aligns with the performance improvements reported in Table 1.